# GENERAL SOLUTION OF THE NAVIER-STOKES SYSTEM, AS PARAMETER FUNCTION AND IT'S GEOPHYSIC AND GYDROAERODINAMIC PRACTICE.

Aleksandr Fridrikson and Marina Kasatochkina

**Abstract** We consider the Navier-Stokes system solution, based at parametric representation of desired function. This solution is unique and it show the velocity of a stream element as its density structure $[\rho_S(x,y,z,t); \vec{\rho}_L(x,y,z,t)]$ function. The solution is smooth, defining conditions of turbulence occurrence and supersonic shock waves in the viscous substance.

Given solution brings to light the genesis of earthquakes and volcanic activity, of turbulence phenomenon and possible stave off natural and anthropogenic catastrophes, connecting with this effect. Given solution allows to evaluate the Bernoulli's law «line density» mechanism for aerodynamical airplanes checkouts and hydromechanical tests correlation, for high-speed extension flood river compensation. In addition given function allows to apply «isostatic breakdown» mechanism for resolution of the Gulf of Mexico and similar problem.

---------------------------------------------------------------------------------------------------------------------

# ОБЩЕЕ РЕШЕНИЕ NAVIER-STOKES SYSTEM, КАК ПАРАМЕТРИЧЕСКАЯ ФУНКЦИЯ И ЕЁ ПРИМЕНЕНИЕ В ГЕОФИЗИКЕ И ГИДРОАЭРОДИНАМИКЕ

Александр Фридрихсон и Марина Касаточкина

**Аннотация**: Рассмотрено общее решение Navier-Stokes system, основанное на параметрическом представлении искомой функции. Результат актуален для понимания единого генезиса процессов сейсмической и вулканической активности, механизма природных и техногенных катастроф, связанных с явлением турбулентности.

Полученное решение выявляет механизм эффекта Бернулли, что весьма важно для более точного проведения аэродинамических и гидромеханических тестов, а также позволяет разработать эффективные способы борьбы с наводнениями на основе высокоскоростной компенсации увеличения расхода в руслах рек. Кроме того, полученная функция даёт основу применить механизм «изостатического пробоя» для высокоскоростной откачки нефти из водного массива в ситуациях подобной катастрофе Мексиканском заливе.

## ВВЕДЕНИЕ

При решении, мы традиционно рассматривали уравнение движения Навье-Стокса, как запись Второго закона Ньютона применительно к индивидуальную частицу $dM = \rho dV$ несжимаемой вязкой среды:

$$\rho \frac{d\vec{u}}{dt} = \frac{d}{dV} \sum \vec{f}$$

где $\sum \vec{f}$ – результирующая всех сил, действующих на индивидуальную частицу.

Поэтому, прежде всего, мы попытались снять проблему «положительной вязкости» в уравнении движения Navier-Stokes system:

$$\frac{d\vec{u}}{dt} = (\vec{F} - \frac{1}{\rho}\text{grad}P) + \frac{\mu}{\rho}\Delta\vec{u}$$

Проблема состоит в следующем: с одной стороны, векторная форма уравнения действительно предусматривает сложение априори противоположно направленных векторов активных и реактивных сил; с другой стороны «положительность» $\mu\Delta\vec{u}$ означает прямую зависимость результирующего ускорения от сил вязкого трения:

$$\mu \to max \ \cup \ \left|\frac{d\vec{u}}{dt}\right| \to max; \ \mu \to min \ \cup \ \left|\frac{d\vec{u}}{dt}\right| \to min$$

при прочих равных условиях - что возможно лишь в случае торможения движущейся по инерции вязкой среды: $\vec{F} = 0 \ \cup \ \frac{d\vec{u}}{dt} < 0$. Однако, как известно, Navier-Stokes system описывает любое движения вязкой среды и прежде всего, под действием не равных нулю массовых сил.

Данное противоречие устраняется преобразованием: ($\Delta\vec{u} = \text{grad div}\,\vec{u} - \text{rot rot}\,\vec{u}$), которое и было применено в (3-1), сделав использование вектора (rot rot $\vec{u}$) физически обоснованным. Ещё более важный аспект состоит в том, что мы изначально не решали задачу Коши и не применяли к решению подход Эйлера:

$$u_i = f(x_i, t)$$

а взяли за основу подход Лагранжа:

$$x_i = f(a_m)$$

где: $a_m$ – физические характеристики движущейся среды;
 $x_i$ – координаты точки их определения

и стали искать общее решение Navier-Stokes system, как функцию вида:

$$u_i = \frac{\partial[x_i(a_m)]}{\partial t}$$

Таким образом, наша задача свелась к тому, чтобы найти такую параметрически заданную функцию $u_i = f[a_m(x, y, z, t)]$, которая удовлетворяла бы уравнениям Navier-Stokes system и хотя бы одному из проверочных критериев Ч.Л. Фефферсмана. Данные критерии были предложены [3] - как математический инструмент проверки предлагаемых общих решений на соответствие их физическому смыслу.

Решая поставленную задачу в векторной форме, мы применили в качестве элементарного вектора «направленный элемент поверхности» $d\vec{S}$ [1] - характеризующий сечение векторной (токовой) трубки и приняли в качестве способа линеаризации нелинейной задачи [2] переход к движущейся с объектом физического исследования неинерциальной - в формализме механики сплошных сред - «сопутствующей системе отсчета» или ССО.

Мы определенным образом ориентировали (*Определение 1*) и связали ССО с произвольно выбранной индивидуальной частицей. Это позволило выявить первый параметр искомого общего решения $a_1 = \vec{\rho}_L \left[\frac{\text{кг}}{\text{м}}\right]$ - «линейную плотность движущейся среды» и привести

Navier-Stokes system к дифференциальному уравнению, выражающему закон сохранения энергии для элементарного объема движущейся вязкой среды (9-3).
Полученное уравнение, в скалярной форме, имеет гладкое интегральное решение, которое, однако, не удовлетворяет проверочному А-критерию Ч.Л.Феффермана [3] и теряет физический смысл при вязком торможении потока.

Продолжив решение полученного уравнения без интегрирования, мы выявили еще один параметр $a_2 = \rho_S$, движущейся среды, связанный со скоростью её движения, который мы условно назвали «поверхностная плотность поперечного сечения потока». Мы объединили обе плотностные характеристики термином «плотностная структура движущейся среды» и выявили их функциональную связь через объемную плотность вещества $\rho$ (*Теорема* 3), которая, в силу уравнения неразрывности Navier-Stokes system, остаётся постоянной при изменении скорости её движения.

Решение структурировано следующим образом.
В части I Navier-Stokes system приводится к уравнению, соответствующему Второму закону Ньютона для индивидуальной частицы, а скорость определяется, как параметрическая функция физических свойств данной жидкой частицы.
В части II полученное уравнение рассматривается в специально ориентированной ССО $[\overrightarrow{dS}_{\text{fgh}}\, dL]$ и преобразуется в дифференциальное уравнение, выражающее закон сохранения энергии для индивидуальной частицы движущейся вязкой среды.
В части III полученное уравнение решается без интегрирования и определяется векторное значение скорости индивидуальной частицы, как функцию параметров, зависящих от времени и координат:

Мы не применяли терминологию и стиль изложения «чистых» математиков в силу большой сложности материала и желания донести его максимально просто. Мы исходили из энциклопедического определения математической формулы: «...от лат. formula — уменьшительное от forma — образ, вид — принятая в математике, а также физике и прикладных науках, символическая запись законченного логического суждения».

Поскольку лишь перечень работ, посвященных решениям Navier-Stokes system, занял бы сотни страниц, в завершении статьи мы указали только те труды, которые были использованы или послужили основой нашей работы. Приносим извинения читателям, изучающим научные статьи с раздела «используемая литература», выясняя чью научную школу представляют авторы. Отвечаем данному читателю: ничью. И нам остается лишь выразить признательность Чарльзу-Льюису Фефферману за его работу [3], в которой дан обзор проблемы Navier-Stokes system по существу.
А также привести цитату из неё, которой мы бы хотели закончить вводную часть:
«Для решения изложенной задачи поиск единственного выровненного решения
Navier-Stokes system - стандартные методы PDE представляются неадекватными.
Вероятно, взамен существующих, нам необходимы новые научные идеи.»

I

Запишем уравнения Navier-Stokes system в векторной форме:

$$（1）\quad \rho\frac{d\vec{u}}{dt} = \rho\left[\frac{\partial \vec{u}}{\partial t} + (\vec{u}\cdot \text{grad})\vec{u}\right] = \rho\vec{F} - \text{grad}P + \mu\Delta\vec{u} \qquad \left[\frac{\text{кг}}{\text{с}^2\,\text{м}^2}\right]$$

$$（2）\quad \text{div}\,\vec{u} = 0 \qquad [\text{с}^{-1}]$$

где: $\vec{u} = \sum_{i=1}^{3} u_i$ - трёхмерный вектор скорости;
$\vec{F}$ - вектор массовых сил, приложенных к вязкой среде;
$P$ - давление в точке определения скорости;
$\rho$ - плотность вещества движущейся среды;
$\mu$ - динамическая вязкость.

Зададим начальные условия:
$$u_i|_{t=0} = u_i^0$$

и определим граничную поверхность $S_0$ движущейся среды, где выполняются условия:
$$u_i|_{S_0} = u_0 \geq 0; \quad P|_{S_0} = P_0$$

*Теорема 1*
Пусть $\sum_{j=1}^{3}(u_i\, u_j) = U_P$ – потенциал поля упругих напряжений движущейся среды. Тогда нелинейное слагаемое левой части уравнения движения Navier-Stokes system преобразуется к виду:
$$(\vec{u}\cdot \text{grad})\vec{u} = \text{grad}\, U_P - \vec{u}\,(\text{div}\,\vec{u})$$

*Доказательство*
Запишем нелинейное слагаемое левой части уравнения движения Navier-Stokes system в скалярной форме и преобразуем в соответствии с правилом дифференцирования произведения:
$$(\vec{u}\cdot \text{grad})\vec{u} = \sum_{j=1}^{3}\frac{\partial u_i}{\partial x_j}u_j = \sum_{j=1}^{3}\left[\frac{\partial(u_i u_j)}{\partial x_j} - \frac{\partial u_j}{\partial x_j}u_i\right]$$

Перепишем полученный результат в векторной форме:

$$\sum_{j=1}^{3}\frac{\partial U_P}{\partial x_j} - \sum_{i=1}^{3}u_i\sum_{j=1}^{3}\frac{\partial u_j}{\partial x_j} = \text{grad}\,U_P - \vec{u}\,(\text{div}\,\vec{u}) = (\vec{u}\cdot\text{grad})\vec{u}$$

Ч.Т.Д.

*Примечание 1*
Если закон сохранения энергии для элементарной массы (индивидуальной частицы) движущейся вязкой среды имеет вид:
$$U_0 = U_{\text{grad}\,P} + U_F + U_P$$

где: $U_{\text{grad}\,P}$ – потенциал поля $\text{grad}P$;
$U_F$ — потенциал поля массовых сил;
$U_P$ — потенциал поля упругих напряжений;
$U_0$ — суммарный потенциал;

и *Теорема 1* верна, то суммарный потенциал в произвольной точке движущейся среды может определяться, как максимальная кинетическая энергия индивидуальной частицы, движущейся относительно трёх плоскостей вязкого трения:

$$U_0 \; = \; \frac{|u_1 + u_2|^2 + |u_2 + u_3|^2 + |u_3 + u_1|^2}{2}$$

Покажем, что такое суждение физически обосновано.
Пусть $u_i$ - скорость жидкой частицы внутри слоя $S_j$ трехмерного течения $X_i$
Рассмотрим такое течение в системе отсчёта, связанной с границей этого течения $S_0$.
Потенциал поля $\text{grad}P$ между слоями $S_j$ и $S_0$ - в соответствии с уравнением Бернулли - может быть выражен интегралом вида:

$$U_{\text{grad }P} \; = \; -\frac{1}{\rho}\int_0^{x_i}\text{grad}P dx \; = \; -\int_{P_0}^{P_j}\frac{dP}{\rho} \; = \; -\int_{u_0}^{u_i}\frac{d}{\rho}\left(P_0 - \frac{\rho u^2}{2}\right) \; = \; \frac{1}{2}(u_i^2 \; - \; u_0^2)$$

В силу закона сохранения, потенциальная энергия поля массовых сил переходит в кинетическую энергию движущейся среды. Это означает, что потенциал поля $i-$ой составляющей массовых сил, может быть выражен неопределенным интегралом вида:

$$U_{F_i} \; = \; \int u_i du \; = \; \frac{u_i^2}{2} \; + \; const \; = \; \frac{1}{2}(u_i^2 \; + \; u_0^2)$$

Тогда сумма двух рассмотренных потенциалов определяется просто:

$$U_{F_i} \; + \; U_{\text{grad }P} = \frac{1}{2}(u_i^2 \; - \; u_0^2) \; + \; \frac{1}{2}(u_i^2 \; + \; u_0^2) \; = \; u_i^2$$

Для трехмерно пространства:

$$\sum_{i=1}^{3} u_i^2 \; = \; u_1^2 \; + \; u_2^2 \; + \; u_3^2 \; = \; U_F \; + \; U_{\text{grad }P}$$

При этом:

(2 − 1) $$\sum_{j=1}^{3}(u_i u_j) \; = \; u_1 u_2 \; + \; u_2 u_3 \; + \; u_3 u_1$$

(2 − 2) $$u_1 u_2 \; = \; \frac{(u_1 + u_2)^2 - (u_1^2 + u_2^2)}{2}$$

В силу эквивалентности преобразования (2-2) для остальных слагаемых (2-1):

(2 − 3) $$\sum_{j=1}^{3}(u_i u_j) \; + \; (u_1^2 + u_2^2 + u_3^2) \; = \; \frac{(u_1 + u_2)^2 + (u_2 + u_3)^2 + (u_3 + u_1)^2}{2}$$

Правая часть (2-3) представляет собой не что иное, как кинетическую энергию элементарной массы вязкой среды (относительно трёх плоскостей вязкого трения), которая, по нашему предположению, определяет значение суммарного потенциала $U_0$. Второе слагаемое левой части (2-3), как мы уже показали:

$$(2-4) \qquad u_1{}^2 + u_2{}^2 + u_3{}^2 = U_F + U_{\text{grad } P}$$

Таким образом:

$$(2-5) \qquad \sum_{j=1}^{3}(u_i u_j) = U_0 - (U_F + U_{\text{grad } P}) = U_P$$

$$(2-6) \qquad U_0 = U_F + U_{\text{grad } P} + U_P$$

Из (2-6) также следует, что для левой части (1), с учётом (2), будет справедливо:

$$(2-7) \qquad \rho\frac{d\vec{u}}{dt} = \rho\left(\frac{\partial \vec{u}}{\partial t} + \text{grad } U_P\right)$$

Выкладки данного примечания приведены в качестве физического обоснования математического условия $\sum_{j=1}^{3}(u_i u_j) = U_P$ и носят иллюстративный характер. При этом, вытекающий из (2-7) «невихревой характер» разности полной и частной производных вектора скорости по времени:

$$(2-8) \qquad \frac{d\vec{u}}{dt} - \frac{\partial \vec{u}}{\partial t} = \text{grad} U_P$$

помогает пониманию физического смысла важнейшего для решения задачи перехода (3-3-2) и (3-3-3). Уравнение (2-8) мы условно назвали уравнением «векторной линии». Оно удовлетворяет движению индивидуальной частицы и, наряду с (2) - может полагаться условием неразрывности векторных линий в ламинарных течениях.

\*\*\*

Теперь, с учётом Теоремы 1, Navier-Stokes system приводится к виду:

$$(3-1) \quad \rho\left[\frac{\partial \vec{u}}{\partial t} + \text{grad} U_P - \vec{u}\,(\text{div }\vec{u})\right] = \rho\vec{F} - \text{grad} P + \mu[\text{grad}\,(\text{div}\,\vec{u}) - (\text{rot rot}\,\vec{u})]$$

$$(3-2) \qquad \text{div }\vec{u} = 0$$

Данная система уравнений, очевидно, сводится к одному уравнению:

$$(3-3-1) \qquad \rho\left(\frac{\partial \vec{u}}{\partial t} + \text{grad } U_P\right) = \rho\vec{F} - \text{grad} P - \mu\,(\text{rot rot}\,\vec{u}) = \rho\frac{d\vec{u}}{dt}$$

Пусть пространство несжимаемой вязкой среды $X_j$ движется в неподвижном пространстве $X_i$ так, чтобы выполнялось условие:

$$t = const \cup X_j \equiv X_i$$

Полагая поле инерционных сил в элементарном объёме движущейся вязкой среды:

$$(3-3-2) \quad -\vec{w}(x_j, t) = \left[\frac{\partial \vec{u}(x_i = const; t)}{\partial t} + \text{grad}U_P(x_j; t = const)\right] = \frac{d\vec{u}}{dt}(x_i; t)$$

и, применив принцип Даламбера, запишем (3-3-1) для неинерциальной системы отсчета, связанной с произвольной индивидуальной частицей, движущейся по заданной векторной линии:

$$(3-3-3) \quad \rho \frac{d}{dt}\vec{u}[a_m(x_i, t)] = \rho\vec{F} - \text{grad}P - \mu(\text{rot rot}\vec{u})$$

где: $a_m(x_i, t)$ - физические характеристики движущейся среды, меняющиеся под действием поля $[-\vec{w}(x_j, t)]$.

Важно отметить, что именно в такой записи уравнение движения принимает вид Второго закона Ньютона - поскольку вязкость находится в обратной зависимости с результирующим ускорением.

<div align="center">II</div>

Выделим элементарный объём токовой (векторной) трубки $dSdL \ll LdS$, который, в определении механики сплошных сред будем полагать индивидуальной частицей.

*Определение 1*

Выберем в качестве сопутствующей системы отсчёта $FGH$ - $\overrightarrow{dS}dL$;
$\overrightarrow{dS}$ **-** принятый в векторном анализе [1] «направленный элемент поверхности» - равный по модулю элементарной площади $dS$ и направленный по нормали к ней;
$dL$ **-** элемент векторной линии, проходящей через центр $\overrightarrow{dS}$, в пределах которого:

$$\frac{\partial u_L}{\partial L} = const;$$

и при $m > 1$, выполняется условие:

$$(3-4) \quad \frac{\partial^m u_L}{\partial L^m} = 0$$

Определим векторную линию $L_0$, для всех точек которой выполняется: $u_L = u_{max}$, термином: «линия максимальной скорости течения». Введём понятие «линия равных скоростей в сечении потока» $l \in S$, для всех точек которой: $u_L = const$ и выполняется условие:

$$(3-5) \quad \frac{\partial^m u_L}{\partial l^m} = 0$$

Пусть $FGH$ движется относительно неподвижной системы отсчёта $XYZ$ в произвольном интервале времени, в произвольно выбранной точке течения.
Введём для $FGH$ ортогональные оси f, g, h и примем следующие условия:
1. **f, g, h** - орты соответствующих осей $FGH$;
2. Ось f направлена по вектору $\vec{F} - \frac{1}{\rho}\text{grad}P = \vec{F}\cos(\vec{F}d\vec{S})$ и, таким образом, по векторной линии $L$, так что выполняется: $\vec{F} - \frac{1}{\rho}\text{grad}P = F_L\mathbf{f}$;
3. Ось h –направлена по вектору $(\text{grad }u_L)$ и определяет ориентацию $FGH$ на $L_0$;
4. $r_\text{h}$ - не обязательно прямая линия, соединяющая $L_0$ и начало координат $FGH$: [f = 0; g = 0; h = 0], при этом: $dh = -dr_\text{h}$;
5. Ось g - ортогональна осям f и h и направлена по касательной к линии $l$ равных скоростей в сечении потока $S$, так что выполняется условие:

$$(3-6) \qquad \partial^m \frac{|u_\text{f} + u_\text{g} + u_\text{h}|}{\partial g^m} = 0$$

6. $FGH$ - сохраняет ориентацию, показанную в пунктах 2,3,4,5, поскольку имеет возможность трёхмерного поворота с угловыми скоростями: $\omega_\text{f} + \omega_\text{g} + \omega_\text{h} = \vec{\omega}$;

7. $\omega_\text{f}$ определяет «винтовое» вращение индивидуальной частицы или поворот $FGH$; $\omega_\text{g}$ и $\omega_\text{h}$ определяют приращения поступательной скорости индивидуальной частицы если её векторная линия $L$ пересекает $d\vec{S}$ не в начале координат (g ≠ 0; h ≠ 0) $FGH$, за счет сдвиговой (динамической) вязкости - так что в общем виде:

$$u_L = u_\text{f} + \omega_\text{h}\text{g} + \omega_\text{g}\text{h}$$

Введём обозначение «ориентированной» ССО $FGH$: $[\vec{dS}_\text{fgh}\, dL]$.

*Примечание 2*
Для наблюдателя, движущегося вместе с вязким течением в системе отсчёта $[\vec{dS}_\text{fgh}\, dL]$ – этот процесс выражается противоположно направленным движением вещества среды по токовым линиям $L_{(\text{g}=0;\, h=+1)}$ и $L_{(\text{g}=0;\, h=-1)}$ со скоростями:

$$+\left[u_{L_{(\text{g}=0;\, \text{h}=+1)}} - u_{L_{(\text{g}=0;\, \text{h}=0)}}\right]; \quad -\left[u_{L_{(\text{g}=0;\, \text{h}=0)}} - u_{L_{(\text{g}=0;\, \text{h}=-1)}}\right]$$

При этом, из физического опыта известно, что такое «встречное» движение иногда действительно образует вихри. Например, для неподвижного наблюдателя, стоящего на берегу реки - такие вихри, на её поверхности, перемещаются со скоростью течения. Для наблюдателя в ССО $[\vec{dS}_\text{fgh}\, dL]$ данное вихревое движение выражается через угловую скорость $\omega_\text{g}$ - парой соответствующих векторов: $\pm\, \omega_\text{g}\text{h}^0\mathbf{f}$

*Определение 2:*
В силу (3-4), (3-6), а также пункта 1 *Определения 1* для $[\vec{dS}_\text{fgh}\, dL]$ справедливо:

$$\frac{\partial^2 u_f}{\partial f^2} = \frac{\partial^2 u_g}{\partial f^2} = \frac{\partial^2 u_h}{\partial f^2} = 0$$

$$\frac{\partial^2 u_f}{\partial g^2} = \frac{\partial^2 u_g}{\partial g^2} = \frac{\partial^2 u_h}{\partial g^2} = 0$$

\*\*\*

Рассмотрим движение несжимаемой вязкой среды в системе отсчёта $\left[\overrightarrow{dS}_{\text{fgh}}\, dL\right]$ и выразим «вихревое слагаемое» $\text{rot}\,\text{rot}\,\vec{u}$ через проекции по осям f, g, h:

$$(4-0) \qquad \text{rot}\,\text{rot}\,\vec{u} \;=\; \text{rot}_f(\text{rot}\,\vec{u}) + \text{rot}_g(\text{rot}\,\vec{u}) + \text{rot}_h(\text{rot}\,\vec{u})$$

$$\text{rot}_f(\text{rot}\,\vec{u}) = \frac{\partial}{\partial g}\text{rot}_h\vec{u} - \frac{\partial}{\partial h}\text{rot}_g\vec{u} = \frac{\partial}{\partial g}\left(\frac{\partial u_g}{\partial f} - \frac{\partial u_f}{\partial g}\right) - \frac{\partial}{\partial h}\left(\frac{\partial u_f}{\partial h} - \frac{\partial u_h}{\partial f}\right) =$$

$$= \frac{\partial^2 u_g}{\partial g \partial f} - \frac{\partial^2 u_f}{\partial g^2} - \frac{\partial^2 u_f}{\partial h^2} + \frac{\partial^2 u_h}{\partial h \partial f} + \frac{\partial^2 u_f}{\partial f^2} - \frac{\partial^2 u_f}{\partial f^2} =$$

$$= \frac{\partial}{\partial f}\left(\frac{\partial u_f}{\partial f} + \frac{\partial u_g}{\partial g} + \frac{\partial u_h}{\partial h}\right) - \left(\frac{\partial^2 u_f}{\partial f^2} + \frac{\partial^2 u_f}{\partial g^2} + \frac{\partial^2 u_f}{\partial h^2}\right)$$

Теперь, с учётом (3-2) и независимости преобразования (3-1) от выбора осей, а также с учётом эквивалентности преобразования $\text{rot}_f(\text{rot}\,\vec{u})$ для оставшихся проекций $\text{rot}_g(\text{rot}\,\vec{u})$ и $\text{rot}_h(\text{rot}\,\vec{u})$ - окончательно получаем:

$$(4-1) \qquad \text{rot}_f(\text{rot}\,\vec{u}) = \frac{\partial}{\partial f}\text{div}\vec{u} - \text{div}\,\text{grad}\, u_f = -\left(\frac{\partial^2 u_f}{\partial f^2} + \frac{\partial^2 u_f}{\partial g^2} + \frac{\partial^2 u_f}{\partial h^2}\right)$$

$$(4-2) \qquad \text{rot}_g(\text{rot}\,\vec{u}) = \frac{\partial}{\partial g}\text{div}\vec{u} - \text{div}\,\text{grad}\, u_g = -\left(\frac{\partial^2 u_g}{\partial g^2} + \frac{\partial^2 u_g}{\partial f^2} + \frac{\partial^2 u_g}{\partial h^2}\right)$$

$$(4-3) \qquad \text{rot}_h(\text{rot}\,\vec{u}) = \frac{\partial}{\partial h}\text{div}\vec{u} - \text{div}\,\text{grad}\, u_h = -\left(\frac{\partial^2 u_h}{\partial h^2} + \frac{\partial^2 u_h}{\partial g^2} + \frac{\partial^2 u_h}{\partial f^2}\right)$$

*Примечание 3*
В неподвижной (инерциальной) системе координат $XYZ$ подобные (4-1),(4-2),(4-3) выражения - можно получить гораздо более простым преобразованием:

$$\Delta\vec{u} = \sum_{i=1}^{3} \text{div}\,\text{grad}\, u_i$$

Однако, в этом случае проблема «положительной вязкости» не решается и (1) не согласуется со Вторым законом Ньютона для $\vec{F} \neq 0$.

\*\*\*

В силу *Определения 2,* сократим полученные выражения (4-1),(4-2),(4-3):

$$(5-1) \quad \operatorname{rot}_f(\operatorname{rot}\vec{u}) = -\frac{\partial^2 u_f}{\partial h^2}$$

$$(5-2) \quad \operatorname{rot}_g(\operatorname{rot}\vec{u}) = -\frac{\partial^2 u_g}{\partial h^2}$$

$$(5-3) \quad \operatorname{rot}_h(\operatorname{rot}\vec{u}) = -\frac{\partial^2 u_h}{\partial h^2}$$

Подставим (5-1), (5-2), (5-3) в (4-0):

$$(6-1) \quad \operatorname{rot}\operatorname{rot}\vec{u} = \operatorname{rot}_f(\operatorname{rot}\vec{u}) + \operatorname{rot}_g(\operatorname{rot}\vec{u}) + \operatorname{rot}_h(\operatorname{rot}\vec{u}) = -\frac{\partial^2}{\partial h^2}u_f - \frac{\partial^2}{\partial h^2}(u_g + u_h)$$

В силу *Определения* 1, преобразуем полученного выражения:

$$(6-2) \quad \operatorname{rot}\operatorname{rot}\vec{u} = -\frac{\partial^2}{\partial r_h^2}u_L + \left[\frac{\partial^2}{\partial h^2}(\omega_g h) + \frac{\partial^2}{\partial h^2}(\omega_h g)\right] - \left[\frac{\partial^2}{\partial h^2}(\omega_f g) + \frac{\partial^2}{\partial h^2}(\omega_f h)\right]$$

Рассмотрим первое слагаемое (6-2). Для вязкого течения существует экспериментально подтверждённое уравнение Пуазейля, которое мы вправе применить для течений ограниченных поверхностью трения, в силу пунктов 3,4 и 5 *Определения 1*, а также при $\vec{F} = f(R^2 - r_h^2)$, где $R$ – условный радиус сечения потока, неограниченного поверхностью трения (морские и океанические течения):

$$(6-3) \quad u(r) = \frac{P_L}{4\mu L}(R^2 - r^2)$$

где: $R$ — расстояние от $L_0$ ($u_L = max$) до ближайшей граничной поверхности;
$r$ — расстояние от $L_0$ до точки определения скорости движущейся среды;
$L$ — протяжённость токовых линий; $P_L$ – перепад давления на участке потока $L$.

Продифференцировав уравнение Пуазейля, выразим первое слагаемое (6-2):

$$(7-1) \quad -\frac{\partial^2 u_L}{\partial r_h^2} = -\frac{d^2 u(r)}{dr^2} = -\frac{P_L}{4\mu L}d^2\frac{(R^2 - r^2)}{dr^2} = \frac{P_L}{2\mu L}$$

*Примечание 4*

В соответствии с теоремой Коши-Гельмгольца, ротор скорости определяет так называемую «завихрённость»: $2\vec{\omega} = \operatorname{rot}\vec{u}$, где: $|\vec{\omega}| = 2\pi/T_{\vec{u}}$; $T_{\vec{u}}$ - период вращения вектора $\vec{u}$. В системе отсчёта $[\overrightarrow{dS}_{fgh}\, dL]$, следствием теоремы Коши-Гельмгольца для плоскости fh (*FOH*) - можно полагать выражение:

$$\left|\frac{\operatorname{rot}\vec{u}}{2}\right| = \omega_g = \frac{u_{L(g=0;h=0)} - u_{L(g=0;h=-1)}}{r_{h-1} - r_h + \delta r} = \frac{u_{L(g=0;h=+1)} - u_{L(g=0;h=0)}}{r_h - r_{h+1} - \delta r} = \frac{2\pi}{T_g}$$

где: $T_g$ - период вращения $\vec{u}$ вокруг оси g;

$r_{h-1} > r_h > r_{h+1}$ в силу пункта 4 *Определения 1*;

$\delta r$ - смещение центра вращения относительно $[f = 0; g = 0; h = 0]$

в силу нелинейности распределения Пуазейля, выражаемого для $[\overrightarrow{dS}_{fgh}\, dL]$:

$$u_{L(g=0;h=0)} - u_{L(g=0;h=-1)} > u_{L(g=0;h=+1)} - u_{L(g=0;h=0)}$$

Это означает, что вектор $\omega_g \mathbf{g}$ смещён относительно начала координат $[\overrightarrow{dS}_{fgh}\, dL]$ по оси h в сторону линии максимальной скорости $L_0$ на величину $\delta h$.

Таким образом, вектор $\text{rot rot }\vec{u}$, в силу пункта 2 *Определения 1*, имеет составляющую, направленную по векторной линии $L$ и, в силу (7-1), зависящую от $P_L$:

$$(7-1-0) \qquad \text{rot rot }\vec{u} \;\geq\; 2\text{rot}(\omega_g \mathbf{g}) \;=\; 4\frac{u_g}{\delta h}\mathbf{f} \;=\; f(P_L)$$

\*\*\*

*Теорема 2*

Рассмотрим течение, описываемое Navier-Stokes system и ограниченное поверхностью $S_0 = 2\pi R L_0$ в цилиндрической системе координат $[z, (r_0 \pm r), \varphi]$, где: $r_0$ – координата линии максимальной скорости течения $L_0(|\vec{u}| = u_{max})$. Пусть произвольно выбранная векторная трубка $L_i \pi dr^2$ внутри границы данного течения может быть разделена (продифференцирована) на «прямые» области $dL_j$, для которых выполняется условие:

$$dL_j = \lim_{r_0 \to \infty}(r_0 \pm r)d\varphi$$

и непрямые, («дугообразные») области $dL_k$, для которых выполняется условие:

$$dL_k = (r_0 \pm r)d\varphi$$

Пусть уравнение Пуазейля справедливо в области течения $dL_j \pi dr^2$.

Требуется доказать:

1. Области $dL_k \pi dr^2$ того же течения удовлетворяет функция вида:

$$\omega_z \;=\; f\frac{(R^2 - r^2)}{r_0 \pm r}$$

где: $\omega_z$ – угловая скорость индивидуальной частицы относительно центра $C_0 \in z$ гипотетической окружности $2\pi(r_0 \pm r)$, которой принадлежит $dL_k$.

2. Уравнению $\text{rot}_z \vec{u} \neq 0$ (при $z \cap S_0$; $-R \leq r \leq R$) удовлетворяют функции:

$$f_1 \;=\; \frac{\partial^2 \omega_z}{\partial r^2} \;=\; k_\mu \frac{R^2}{r^3} \text{; при } 0 < r < R$$

$$f_2 \;=\; \frac{\partial^2 \omega_z}{\partial r^2} \;=\; 0\text{; при } 0 > r > -R$$

где: $k_\mu$ – коэффициент пропорциональности, зависящий от сдвиговой вязкости.

Доказательство:

Уравнение (1) Navier-Stokes system в области течения $dL_k \pi dr^2$ определяет условие равновесия немассовых сил, действующих на индивидуальную частицу:

$$\omega_z^2 (r_0 \pm r) - \frac{1}{\rho} \text{grad}_r P = 0$$

Тогда, в силу уравнения (2) Navier-Stokes system и закона сохранения импульса индивидуальной частицы течения, находящейся только под действием массовых и «вязких» сил - при переходе из области $dL_j \pi dr^2$ в область $dL_k \pi dr^2$ - выполняется:

$$\frac{\partial L_j}{\partial t} = \frac{\partial L_k}{\partial t}$$

Таким образом, если уравнение Пуазейля справедливо в пределах $dL_j \pi dr^2$ и для всех $r \leq R = const$:

$$u_L = f(R^2 - r^2) = \lim_{r_0 \to \infty} [\omega_z (r_0 \pm r)]$$

то в пределах $dL_k \pi dr^2$, для всех $r \leq R = const$, справедлива функция:

$$\omega_z = \frac{u_L}{(r_0 \pm r)} = f \frac{(R^2 - r^2)}{r_0 \pm r}$$

Положение 1 теоремы доказано.
Вихрю $\text{rot}_z \vec{u}$ соответствуют две асимптоты полученной функции:

$$\lim_{r_0 \to 0} \left[ \frac{(R^2 - r^2)}{r_0 + r} \right] = f(R^2 r^{-1} - r)$$

$$\lim_{r_0 \to R} \left[ \frac{(R^2 - r^2)}{r_0 - r} \right] = f \left[ \frac{(R-r)(R+r)}{R-r} \right] = f(r)$$

Дважды продифференцировав данные функции, окончательно получим:

$$\frac{\partial \omega_z}{\partial r} = -f \left( \frac{R^2}{r^2} \right) - const; \quad \frac{\partial^2 \omega_z}{\partial r^2} = f \left( \frac{R^2}{r^3} \right)$$

$$\frac{\partial \omega_z}{\partial r} = const; \quad \frac{\partial^2 \omega_z}{\partial r^2} = 0$$

Таким образом, вихрю $\text{rot}_z \vec{u}$ удовлетворяют два уравнения, описывающие две области данного вихря:

$$\frac{\partial^2 \omega_z}{\partial r^2} = k_\mu \frac{R^2}{r^3}; \ 0 < r < R$$

$$\frac{\partial^2 \omega_z}{\partial r^2} = 0; \quad 0 > r > -R$$

Положение 2 теоремы доказано.
Ч.Т.Д.

В декартовой системе координат *Теорема 2* может быть сформулирована аналогично:
Если некоторая область ламинарного течения удовлетворяет уравнению Пуазейля и выполняется:

$$\frac{\partial^2}{\partial x_k^2} u_i = const$$

то для $rot_j \vec{u} \neq 0$, того же течения, выполняются условия:

(7 – 1 – 1) $$\frac{\partial^2}{\partial x_k^2} \omega_j = R^2 \frac{k_\mu}{x_k^3} = f_\omega ; \ x_k > x_0$$

(7 – 1 – 2) $$\frac{\partial \omega_j}{\partial x_k} = \partial_{x_k} \frac{rot_j \vec{u}}{2} = const; \quad \frac{\partial^2}{\partial x_k^2} \omega_j = 0; \ \ x_k < x_0$$

где: $R$ – радиус вихря; $x_0$ – определяет границу его внешней и внутренней областей.

*Примечание 5*
Данное разделение областей вихря наблюдаются при круговом движении жидкости, ограниченной поверхностью трения – близ неё, на периферии вихря угловая скорость вращения заметно нарастает при удалении от граничной поверхности; в центральной области вихря – угловая скорость изменяется незначительно – так, будто вязкая среда вращается - как «единое целое».
При образовании вихревой «воронки» - наоборот: центральная часть характеризуется резким ростом угловой скорости ($\partial f_\omega / \partial x_k \to max$); периферия - квазипостоянным значением функции ($\partial f_\omega / \partial x_k \to min$).
Из (7-1-1) вытекает, что для $L_j \pi R^2$, при $(\sum_{k=1}^{2} x_k^2)^{\frac{1}{2}} = r \leq R$ - выполняется предел:

$$\lim_{r \to max \ ; \ R \to min} \left(\frac{R^2}{r^3}\right) = \frac{1}{k_\mu} \frac{\partial^2 \omega_j}{\partial x_k^2} = const$$

\*\*\*

Преобразуем остальные слагаемые (6-2), приняв $\frac{\partial g}{\partial h} = k_{gh} = const \ \cup \ \frac{\partial^2 g}{\partial h^2} = 0$:

$$\partial_h \left[\frac{\partial(\omega_g h)}{\partial h}\right] = \partial_h \left(\omega_g \frac{\partial h}{\partial h} + h \frac{\partial \omega_g}{\partial h}\right) = \frac{\partial \omega_g}{\partial h} + \frac{\partial h}{\partial h}\frac{\partial \omega_g}{\partial h} + h\frac{\partial^2 \omega_g}{\partial h^2} = 2\frac{\partial \omega_g}{\partial h} + h\frac{\partial^2 \omega_g}{\partial h^2}$$

$$\partial_h \left[\frac{\partial(\omega_h g)}{\partial h}\right] = \partial_h \left(\omega_h \frac{\partial g}{\partial h} + g \frac{\partial \omega_h}{\partial h}\right) = \frac{\partial \omega_h}{\partial h}\frac{\partial g}{\partial h} + \frac{\partial g}{\partial h}\frac{\partial \omega_h}{\partial h} + g\frac{\partial^2 \omega_h}{\partial h^2} = 2\frac{\partial \omega_h}{\partial h}\frac{\partial g}{\partial h} + g\frac{\partial^2 \omega_h}{\partial h^2}$$

$$\left[\frac{\partial^2}{\partial h^2}(\omega_g h) + \frac{\partial^2}{\partial h^2}(\omega_h g)\right] = 2k_{gh}\frac{\partial \omega_h}{\partial h} + 2\frac{\partial \omega_g}{\partial h} + \frac{\partial^2 \omega_h}{\partial h^2}g + \frac{\partial^2 \omega_g}{\partial h^2}h$$

$$-\left[\frac{\partial^2(\omega_f h)}{\partial h^2} + \frac{\partial^2(\omega_f g)}{\partial h^2}\right] = -2k_{gh}\frac{\partial \omega_f}{\partial h} - 2\frac{\partial \omega_f}{\partial h} - \frac{\partial^2 \omega_f}{\partial h^2}g - \frac{\partial^2 \omega_f}{\partial h^2}h$$

14Заменив, в силу *Определения* 1, $d\mathrm{h} = -dr_h$ , с учётом Теоремы Коши-Гельмгольца и (7-1-0) - получим окончательное выражение исследуемого вектора:

$$(7-2) \qquad \mathrm{rot}\,\mathrm{rot}\,\vec{\mathrm{u}} =$$

$$\left[\frac{\partial \mathrm{rot}_f \vec{\mathrm{u}}}{\partial r_h}(k_{\mathrm{gh}}+1) + \frac{\partial \mathrm{rot}_g \vec{\mathrm{u}}}{\partial \mathrm{h}} + k_{\mathrm{gh}}\frac{\partial \mathrm{rot}_h \vec{\mathrm{u}}}{\partial \mathrm{h}}\right] + \frac{P_L}{2\mu L}\mathbf{f} + \left[\left(\frac{\partial^2 \omega_\mathrm{h}}{\partial \mathrm{h}^2}\mathbf{g} + \frac{\partial^2 \omega_\mathrm{g}}{\partial \mathrm{h}^2}\mathbf{h}\right) - \frac{\partial^2 \omega_\mathrm{f}}{\partial \mathrm{h}^2}(\mathbf{g}+\mathbf{h})\right]$$

В силу (7-1-2) первое слагаемое полученного выражения - есть постоянная величина, определяющая способность движущейся среды к образованию «глобального» вихря вокруг $L_0$ , а также - локальных вихрей, за счет вязкого трения между смежными векторными линиями течения.

Второе слагаемое, при $P_L(t) = const$ – также величина постоянная. Данный параметр, определяет стабильность векторных линий течения в плоскости его сечения или, что то же самое - способность индивидуальной частицы сопротивляться энергетически выгодному, при определённых условиях, «винтовому» движению.

Третье слагаемое описывает вихревое движение близ поверхности трения или в вихревых «воронках» (*Примечание* 5) и поэтому, в силу (7-1-2), для неограниченных поверхностью трения течений:

$$\left[\left(\frac{\partial^2 \omega_\mathrm{h}}{\partial \mathrm{h}^2}\mathbf{g} + \frac{\partial^2 \omega_\mathrm{g}}{\partial \mathrm{h}^2}\mathbf{h}\right) - \frac{\partial^2 \omega_\mathrm{f}}{\partial \mathrm{h}^2}(\mathbf{g}+\mathbf{h})\right] = 0$$

Таким образом, выражение (4-0) приобретает окончательный вид:

$$(7-3) \qquad \mathrm{rot}\,\mathrm{rot}\,\vec{\mathrm{u}} = \frac{P_L}{2\mu L}\mathbf{f} + \mu_{rot}[\mathbf{f}+\mathbf{g}+\mathbf{h}] + \mu_\omega[\mathbf{g}+\mathbf{h}] \qquad \left[\frac{1}{\mathrm{м}\cdot\mathrm{с}}\right]$$

где: $\mu_{rot} = \left|\frac{\partial \mathrm{rot}_f \vec{\mathrm{u}}}{\partial r_h}(k_{\mathrm{gh}}+1) + \frac{\partial \mathrm{rot}_g \vec{\mathrm{u}}}{\partial \mathrm{h}} + k_{\mathrm{gh}}\frac{\partial \mathrm{rot}_h \vec{\mathrm{u}}}{\partial \mathrm{h}}\right|$; $\mu_\omega = \left|\left(\frac{\partial^2 \omega_\mathrm{h}}{\partial \mathrm{h}^2}\mathbf{g} + \frac{\partial^2 \omega_\mathrm{g}}{\partial \mathrm{h}^2}\mathbf{h}\right) - \frac{\partial^2 \omega_\mathrm{f}}{\partial \mathrm{h}^2}(\mathbf{g}+\mathbf{h})\right|$

Выражение (7-3) показывает, что вектор $\mu(\mathrm{rot}\,\mathrm{rot}\,\vec{\mathrm{u}})$ - в уравнении (1) - должен быть отрицательным, так как не направлен против скорости движения индивидуальной частицы:

$$(8) \qquad \vec{\mathrm{u}} = u_L \mathbf{f} = \frac{dL}{dt}$$

*Определение 3*
Будем полагать $\vec{\mu}_{rot} = \mu_{rot}[\mathbf{f}+\mathbf{g}+\mathbf{h}] + \mu_\omega[\mathbf{g}+\mathbf{h}]$ и $\vec{\mathfrak{K}} = (P_L/2L)\mathbf{f}$ векторными характеристиками движущейся вязкой среды, обозначив: $\vec{\mu}_{rot}$ - «вихревая вязкость» или «вязкая вихреспособность»; $\vec{\mathfrak{K}}$ - «динамическая вихресопротивляемость».

Преобразуем (3-3-3) с учётом (7-3):

$$(9-1) \qquad \rho\frac{d\vec{\mathrm{u}}}{dt} = \rho\vec{\mathrm{F}} - \mathrm{grad}P - \frac{P_L}{2L}\mathbf{f} - \mu\vec{\mu}_{rot}$$

*Примечание 6*
Вихревая вязкость $\vec{\mu}_{rot}$, по физическому смыслу аналогична, но не эквивалентна так называемой «турбулентной вязкости». Влияние $\vec{К}$ на движение жидкости несложно видеть в таком каждодневном опыте, как спуск воды из ванны. Вихрь возникает, при определенном понижении уровня воды – когда её вихресопротивляемость падает за счёт уменьшения $P_L = \rho g_h \Delta h$. Другой опыт показывает, что удлиняя «носик» заливной воронки - при фиксированном уровне воды в её горловине мы можем наблюдаем возникновение вихря, при снижении вихресопротивляемости, за счёт увеличения $L$.

*Определение 4*
В силу закона сохранения энергии, поле любого источника массовых сил - можно выразить, как функцию перепада давления $P_L$, создаваемого $\vec{F}$ на участке $L$ движущейся среды массой $M$:

$$\vec{F} = \frac{P_L d\vec{S}}{dM}$$

Докажем это положение. Рассмотрим (1) для $\rho\vec{F} - \text{grad}P = 0;\ \vec{F}(t) = const$.
При таких условиях:

$$\frac{d\vec{u}}{dt} = 0 \cup \rho\vec{F} = \text{grad}P = const = \frac{P_L}{L}$$

Отсюда, в силу $dV = LdS$ *Определения* 1, получим искомое выражение:

$$\vec{F} = \frac{1}{\rho}\text{grad}P = \frac{dS}{dM}L\,\text{grad}P = \frac{P_L d\vec{S}}{dM}$$

Введём полученное для $\vec{F}$ выражение в (9-1):

$$(9-2) \qquad \rho\frac{d\vec{u}}{dt} = \rho\frac{P_L d\vec{S}}{dM} - \text{grad}P - \vec{К} - \mu\vec{\mu}_{rot}$$

Умножим обе части (9-2) на $dL$ и учтём (8):

$$\rho\frac{d\vec{u}}{dt}dL = \rho P_L \frac{d\vec{S}}{dM}dL - \partial P\left(\frac{1}{\partial f} + \frac{1}{\partial g} + \frac{1}{\partial h}\right)dL - \vec{К}dL - \mu\vec{\mu}_{rot}dL$$

$$\rho\frac{\overrightarrow{dL}}{dt}du = \rho P_L \frac{dL}{dM}d\vec{S} - \left(\frac{\partial L}{\partial f} + \frac{\partial L}{\partial g} + \frac{\partial L}{\partial h}\right)dP - \vec{К}dL - \mu\vec{\mu}_{rot}dL$$

$$\rho\vec{u}du = \rho P_L \frac{dL}{dM}d\vec{S} - \beth\,dP - (\vec{К} + \mu\vec{\mu}_{rot})dL$$

где: $\beth$ - безразмерный вектор, определяющий направление поля $\text{grad}P$ в данной точке движущейся среды: при обтекании крыла он совпадает с вектором подъемной силы;

при торможении о препятствие или разгоне от лопасти турбины - направлен против или по вектору массовых сил - во всех случаях указывая направление удлинения $L$.

В принятой нами ССО $\left[\overrightarrow{dS}_{\text{fgh}}\, dL\right]$ линейная плотность - есть векторная величина, характеризующая интервал движения индивидуальных частиц по заданному, векторной линией, направлению:

$(9-1)$ $$\lim_{L\to 0}\frac{M}{LdS}d\vec{S} \;=\; \frac{dM}{dL}\frac{d\vec{S}}{dS} \;=\; \vec{\rho}_L \;\; ;$$

$(9-2)$ $$\frac{dL}{dM} \;=\; \frac{1}{|\vec{\rho}_L|}$$

Применив (9-1) и (9-2), получим закон сохранения энергии для индивидуальной частицы движущейся вязкой среды при условии: $P_L(t)=const$.

$(9-3)$ $$\rho\vec{u}du \;+\; \beth dP \;+\; (\vec{\mathfrak{K}} \;+\; \mu\vec{\mu}_{rot})dL \;-\; P_L\frac{\rho}{|\vec{\rho}_L|}d\vec{S} \;=\; 0 \qquad \left[\frac{j}{m^3}\right]$$

Данное уравнение интегрируется в скалярной форме и даёт гладкое решение вида:

$$u_i(x_i;t) \;=\; \left[\frac{2}{\rho}(f_1 \;-\; f_2 \;-\; P)\right]^{\frac{1}{2}}$$

где: $f_1[F_i(x_i;t); |\vec{\rho}|_L(x_i;t)]$ - функция параметров $i$-составляющей вектора массовых сил;
$f_2(\mu; |\vec{\mu}|_{\text{rot}}; |\vec{\mathfrak{K}}|)$ - функция параметров $i$-составляющей вектора реактивных сил;
$P(x_i,t)$ и $\rho=const$ - статическое давление и плотность в точке определения $u_i$.

Очевидно, что данное решение не удовлетворяет проверочному А- критерию Ч.Л.Феффермана [3] и теряет физический смысл при: $f_1=0;\ f_2>0;\ P>0$. Продолжим решение полученного уравнения без интегрирования.

II

Заменим $\vec{\mathfrak{K}} \;=\; (P_L/2L)\mathbf{f} \;=\; \dfrac{\vec{P_L}}{2L}$ :

$$\vec{u}du \;=\; \frac{P_L}{|\vec{\rho}_L|}d\vec{S} \;-\; \frac{\vec{P_L}}{2\rho L}dL \;-\; \frac{\mu}{\rho}\vec{\mu}_{rot}dL \;-\; \beth\frac{dP}{\rho}$$

$$\vec{u} \;=\; \frac{P_L}{|\vec{\rho}_L|}\frac{d\vec{S}}{du} \;-\; \frac{\vec{P_L}}{2\rho L}\frac{dL}{du} \;-\; \vartheta\vec{\mu}_{rot}\frac{dL}{du} \;-\; \frac{\beth}{\rho}\frac{dP}{du}$$

где: $\vartheta \;=\; (\mu/\rho)$ - кинематическая вязкость движущейся среды.

$$\vec{u} = \frac{dL}{du}\left(\frac{P_L}{|\vec{\rho}_L|}\frac{d\vec{S}}{dL} - \frac{\vec{P_L}}{2\rho L} - \vartheta\vec{\mu}_{rot}\right) - \frac{\vec{\beth}}{\rho}\frac{dP}{du}$$

В ССО $[\overrightarrow{dS}_{\text{fgh}}\, dL]$, объемная плотность – не что иное, как плотность векторной трубки:

$$\rho = \frac{dM}{dV} = \frac{dM}{LdS}$$

Тогда:

$$\frac{\vec{P_L}}{2\rho L} = \frac{P_L}{2}\frac{d\vec{S}}{dM}$$

(10) $$\vec{u} = \theta_0\left[P_L\left(\frac{\vec{\theta}_1}{|\vec{\rho}_L|} - \frac{\vec{\theta}_2}{\rho_S}\right) - \vartheta\vec{\mu}_{rot}\right] - \vec{\theta}_3$$

где: $\theta_0 = \frac{dL}{du}$; $\vec{\theta}_1 = \frac{d\vec{S}}{dL}$; $|\vec{\theta}_2| = \frac{1}{2}$, в силу: $\frac{1}{\rho_S} = \frac{d\vec{S}}{dM}$; $\vec{\theta}_3 = \frac{\vec{\beth}}{\rho}\frac{dP}{du}$

*Определение 3*
Введём обозначения плотностных характеристик движущейся вязкой среды:
$\vec{\rho}_L$ - «продольная плотность движущейся среды» (плотность векторной трубки) $\left[\frac{\text{кг}}{\text{м}}\right]$;
$\rho_S$ - «поперечная плотность движущейся среды» (плотность векторных линий) $\left[\frac{\text{кг}}{\text{м}^2}\right]$;
$\vec{\theta}_1$ - вектор, характеризующий способность вязкого потока сужаться по направлению поступательного движения в ламинарном режиме; равно как и обратную способность расширяться по направлению движения, при снятии ограничивающей поверхности. В этом случае, $\vec{\theta}_1$ можно определить, как «растр свободной струи» или «свободу потока» (при этом, $\vec{\theta}_1$ очевидно применим для определения через него параметров $\vec{\rho}_L$ и $\rho_S$);
$\vec{\theta}_2$ - безразмерный вектор, характеризующий направление векторной трубки $Ld\vec{S}$;
$\vec{\theta}_3$ - вектор, характеризующий изменение давления $P$ в пространстве и времени, так как его направление даёт $\text{grad}P(x,y,z)$, а модуль определяет $\frac{dP}{du}(t)$.

*Теорема 2:*
Пусть в трёхмерной среде массой $M = mM_{LdS} = mM_{SdL}$, где $m$ – условно выбранное число векторных трубок $LdS$ и условно равное ему число поперечных слоёв $SdL$, выполняются условия:

$$\vec{u}(x,y,z,t) \neq const;\ \rho_S \neq const;\ \vec{\rho}_L \neq const;\ \rho(x,y,z,t) = const.$$

Требуется доказать, что при данных условиях: $|\rho_S\vec{\rho}_L| = const = f(\rho)$

Доказательство:
1. Пусть линейная плотность векторной трубки данной движущейся среды - есть функция времени и не меняется в пространстве:

$$\vec{\rho}_L(t) = \frac{dM_{LdS}}{dL} = \frac{M_{LdS}}{L} = const$$

По условию: $M_{LdS} = M_{SdL} = M/m$ .
Тогда:

$$\rho = \frac{dM_{LdS}}{LdS} \cup \rho M_{LdS} = \frac{dM_{SdL}}{dS}\frac{M_{LdS}}{L} = \rho_S|\vec{\rho}_L| = const$$

2. Пусть линейная плотность векторной трубки данной движущейся среды в данный момент времени - есть функция координат:

$$\vec{\rho}_L(t = const) = \frac{dM_{LdS}}{dL}(x,y,z) \neq const$$

Тогда:

$$M_{LdS} = M_{SdL} = \int_0^{S_i}(\rho_S)_i dS = \int_0^{L_j}|(\vec{\rho}_L)_j|dL \ ;$$

Однако, при этом:

$$\rho = \frac{dM}{dV} = \frac{dM}{d(S_i L_j)}$$

Отсюда:

$$\frac{1}{\rho} = \frac{S_i dL_j + L_j dS_i}{m dM_{LdS}} = \frac{S_i}{m|(\vec{\rho}_L)_j|} + \frac{L_j}{m(\rho_S)_i} = \frac{(\rho_S)_i S_i + |(\vec{\rho}_L)_j|L_j}{m|(\vec{\rho}_L)_j|(\rho_S)_i} = \frac{M_{SdL} + M_{LdS}}{m|(\rho_S)_i(\vec{\rho}_L)_j|}$$

Таким образом:

$$|(\rho_S)_i(\vec{\rho}_L)_j| = \frac{\rho}{m}(M_{SdL} + M_{LdS}) = \left(\frac{2M}{m^2}\right)\rho = const$$

Ч.Т.Д.

III

Определим $\theta_0$ через известные физические характеристики механики сплошных сред. Запишем уравнение распространения продольной упругой волны через массу $dM$ элементарного объёма среды $LdS$ в интервале времени $dt$:

$$\frac{dE_F}{dM}dt = cdL = \lambda_0 du$$

где: $c$ - скорость звука в данной среде
$E_F$ - энергия распространяемого поля сил упругости;
$\lambda_0$ - длина волны собственных упругих колебаний данной среды;
$u$ - скорость смещения частицы среды, при прохождении волны упругости.

Отсюда:

(11 − 1)  $$\theta_0 = \frac{dL}{du} = \frac{\lambda_0}{c} = \frac{1}{\omega_0}$$

где: $\omega_0$ - частота собственных (свободных) колебаний вещества данной среды.

Проведём преобразование:

(11 – 2) $$\vec{\theta}_3 = \frac{\vec{\beth}}{\rho}\frac{dP}{du} = \frac{\vec{\beth}}{\rho}\left(\frac{dPdL}{dLdu}\right) = \frac{\vec{\beth}}{\rho\omega_0}\left(\frac{dP}{dL}\right)$$

Из (11-2) ясно, что в стационарных течениях, при удалении от источника массовых сил типа турбины, много большем, чем размер её лопастей: $P(L) \approx const \ \cup \ \vec{\theta}_3 \to 0$.

Применив векторную форму (11-2), запишем закон сохранения энергии конечно малой массы $m_L = \rho V_L$ движущейся со скоростью $\vec{u}_0$ - при переходе её бесконечно малой кинетической энергии $m_L|\vec{u}_0|du$ в бесконечно малую упругую энергию $V_L dP$:

$$m_L\vec{u}_0 du = -\left(\frac{m_L}{\rho}dP\right)\vec{\beth}$$

Отсюда:

$$\vec{u}_0 = -\frac{1}{\rho}\frac{dP}{du}\vec{\beth} = -\vec{\theta}_3$$

Таким образом:

(11 – 3) $$\vec{u} = \vec{u}_L + \vec{u}_0$$

где: $\vec{u}_L = \frac{1}{\omega_0}\left[P_L\left(\frac{\vec{\theta}_1}{\rho_L} - \frac{\vec{\theta}_2}{\rho_S}\right) - \vartheta\vec{\mu}_{rot}\right]$

Из (11-3) следует:
$\vec{u}_L$ – есть скорость индивидуальной частицы относительно «своей» векторной трубки;
$\vec{u}_0$ – есть скорость «упругой отдачи» или «смещения» самой векторной трубки относительно источника массовых сил. Таким образом, $\vec{u}$ – есть скорость индивидуальной частицы относительно источника массовых сил. Это означает инвариантность искомой параметрической функции относительно связанной с источником массовых сил неподвижной декартовой и принятой нами систем отсчёта:

$$\vec{u}(f, g, h, t) \equiv \vec{u}(x, y, z, t)$$

Запишем полученное решение Navier-Stokes System в декартовой системе отсчёта:

(12) $$\vec{u}(x, y, z, t) = \frac{1}{\omega_0}\left[P_L\left(\frac{\vec{\theta}_1}{|\vec{\rho}_L|} - \frac{\vec{\theta}_2}{\rho_S}\right) - \vartheta\vec{\mu}_{rot}\right] - \vec{\theta}_3$$

где: $\quad \vec{\rho}_L = f_1(x, y, z, t) \neq 0; \ \rho_S = f_2(x, y, z, t) \neq 0; \ P_L = f_3(t)$

Решение удовлетворяет А-критерию Ч. Л. Феффермана [3] и может полагаться общим:

$$\vartheta > 0; \quad \vec{F} = 0$$

$$\frac{P_L d\vec{S}}{dM} = 0 \ \cup \ P_L = 0; \quad \frac{dP}{du} < 0$$

$$\vec{u} = \frac{\vec{\beth}}{\rho}\frac{dP}{du} - \frac{\vartheta \vec{\mu}_{rot}}{\omega_0} \ ; \quad t_0 = \left[\frac{\vec{\beth}}{\rho(\vartheta\vec{\mu}_{rot})}\right]\frac{dP}{du} - \frac{1}{\omega_0}$$

Данный критерий, применительно к решению (12) – показывает, что в отсутствии поля массовых сил, движущаяся по инерции вязкая среда полностью останавливается за конечное время $t_0$, определяемое скоростью потока на момент «выключения» поля массовых сил и «вязкими» свойствами среды.

## ВЫВОДЫ

Полученное решение является единственным и «гладким», дифференцируясь во всём множестве независимых переменных, кроме:

$$|\vec{\rho}_L| = 0 \ \cup \ |\vec{u}| = +\infty$$

$$\rho_S = 0 \ \cup \ |\vec{u}| = -\infty$$

В первом случае, показанные значения определяют продольный разрыв токовых (векторных) линий – что, очевидно, является условием перехода ламинарного потока в состояние турбулентности; во втором случае – «поперечный разрыв сплошности» и условие возникновения ударной волны в сверхзвуковом режиме движения среды.

При $|(\partial_t\vec{\rho}_L)\rho_S + (\partial_t\rho_S)\vec{\rho}_L| \gg [(\partial_t\rho)M + (\partial_t M)\rho]$, (12) можно полагать справедливым и для сжимаемых вязких сред. Это означает необходимость учета параметров $\vec{\rho}_L$ и $\rho_S$, при испытании, например, летательных аппаратов в аэродинамических трубах, где движение самолёта через неподвижный воздух приравнивается к движению воздуха через неподвижный самолёт. Как показывает общее решение Navier-Stokes system, подобное приравнивание не корректно, в силу изменения плотностной структуры движущейся воздушной среды по сравнению с неподвижным воздухом.

Решение (12), при $P = f(|\vec{\rho}_L|)$ – определяет плотностной механизм эффекта Бернулли:

$$\vec{u} \to max \ \cup \ \rho_S \to max \ \cup \ \vec{\rho}_L \to min \ \cup \ P \to min$$
$$\vec{u} \to min \ \cup \ \rho_S \to min \ \cup \ \vec{\rho}_L \to max \ \cup \ P \to max$$
$$\rho = const$$

## ПРАКТИЧЕСКОЕ ПРИМЕНЕНИЕ

Решение (12) показывает возможность интенсивного падения статического давления в конвекционных магматических потоках на границе жидкого ядра и нижней мантии, где выполняются условия $\rho \to max; \ \vec{u} \to \min; \ \vartheta \to min$ и есть предпосылки:

$$\frac{dP}{du}\vec{\beth} = \left\{\frac{\rho}{\omega_0}\left[P_L\left(\frac{\vec{\theta}_1}{|\vec{\rho}_L|} - \frac{\vec{\theta}_2}{\rho_S}\right) - \vartheta\vec{\mu}_{rot}\right] - \vec{u}\right\} \to \max$$

которые определяют возможный генезис очагов землетрясений и вулканической активности за счёт «барического» вскипания конвекционных расплавов.

Решение (12) даёт основу для точного расчёта и разработки технического способа эффективной откачки загрязняющих веществ непосредственно из водного или воздушного массива, за счёт «высокоскоростного втягивания» при условии:

$$\vec{u}_\text{т} = \frac{1}{\omega_0}\left[P_L\left(\frac{\vec{\theta}_1}{|\vec{\rho}_L|} - \frac{\vec{\theta}_2}{\rho_S}\right) - \vartheta\vec{\mu}_{rot}\right] - \vec{\theta}_3 \geq \sum_{i=1}^{3} u_{3i}$$

где: $\vec{u}_\text{т}$ – скорость вертикального потока в откачивающем трубопроводе;
$u_{3i}$ – скорость выброса загрязняющих веществ.

Для решения проблемы эффективной откачки нефти в Мексиканском заливе необходимы начальные условия: $\vec{u}_\text{т} \approx 10^2 \frac{\text{м}}{\text{сек}}$; $\Delta P \cong \rho g h = 10^7 \text{Па}$, при выполнении которых возникает гидродинамический процесс с положительной обратной связью:

$$\vec{u} \to max \cup P \to min \cup P_L \to max \cup \vec{u} \to max \cup \frac{dP}{du}\vec{\beth} \to max \cup \vec{u}_0 \to max \cup \vec{u} \to max$$

который компенсирует силы вязкого трения и позволяет поддерживать напор малой мощностью (эффект фонтана при забуривании в нефтепласт). Проблема в том, что при аварийном разрыве магистрали - вновь «зацепить» пластовое давление нельзя именно из-за обрыва цепи положительной обратной связи. Разогнать же с нуля до скорости её «запуска» километровый водный столб можно при мощностях порядка 10МВт – что естественно снижает КПД процесса и технически трудноосуществимо.
Не «единственно», но возможным решением является «изостатический пробой» - подобный тому, какой происходит в высокоскоростных восходящих потоках при возникновении торнадо. Технические способы создать ИП близ устья разрушенной нефтескважины в Мексиканском заливе существуют. Авторы и редакция готовы изложить их суть при соответствующем обращении руководства компании BP.

## ПЕРСПЕКТИВЫ

Представляется перспективным поиск связи критического числа Рейнолдса $Re^*$ и «характерной скорости» $u^*$ с параметрами плотностной структуры вязкой среды. Выявлена функция $u^* = f(Re^*, \vec{\rho}_L, M^{-1})$, которая выводит на турбулентный механизм триггера таких явлений, как ураганы и смерчи и может стать теоретической основой разработки способов борьбы с наводнениями посредством высокоскоростной прокачки водного массива рек в пределах населённых пунктов при $\vec{u} < u^*$.

Представляется актуальным поиск условий, при которых возникают и стабильно перемещаются, со скоростью течения, вихри - показанные в *Примечании 2*, а также исследование вопроса: связаны ли подобные процессы с явлением турбулентности.

Представляется необходимым исследование связи применяемых при решении характеристик вязкой среды и ограничений для Navier-Stokes system, показанных в [3]: прежде всего, асимптот $C_{\propto K}(1+|x|)^{-K}$ и $C_{\propto mK}(1+|x|+t)^{-K}$. Данная связь, для вторых производных, в $[\vec{dS}_{fgh}\, dL]$ - может быть выражена пределами:

$$\lim_{x \to 0}\left[\partial_h \frac{\partial u^0(x_g)}{\partial x_h}\right] = |\vec{\mu}_{rot}| \leq C_{22}; \qquad \lim_{x \to 0; t \to 0} \partial_x\, \partial_t |\vec{F}| = \omega_f \omega_g \omega_h \leq C_{111} = \omega_0^3$$

Правое выражение, например, означает - что частота приложения к среде поля массовых сил должна быть не больше частоты собственных колебаний данной среды. Отсюда следует, что Navier-Stokes system не рассматривает «вибрационный» характер воздействия поля массовых сил на вязкую среду - перемещаемую, в таком случае, как единое целое.

\*\*\*

Мы выражаем искреннюю признательность всем, кто работал и работает над проблемой Navier-Stokes system, надеясь продолжить исследования с целью практического применения результатов - в аэродинамике и геофизике.